\documentclass[12pt]{iopart}
\usepackage{graphicx}
\eqnobysec

\begin{document}

\title{A Novel Variational Principle in Electrostatics and its Consequences}

\author{Kolahal Bhattacharya}

\address{C-103, HECR Hall, T.I.F.R. Mumbai}
\ead{kolahalb@tifr.res.in}
\begin{abstract}
We propose a novel variational principle in electrostatics and show that one can derive mirror 
equation in the context of image problem starting from this principle. The corresponding Euler-Lagrange 
equation is seen to lead to Green's differential equation (also known as Thomson's equation).
\end{abstract}

\maketitle

\section{Introduction}
The method of images is a well-known powerful tool for solving boundary value problems in 
electromagnetism. In 1848, Sir W Thomson (also known as Lord Kelvin) introduced this method in a paper 
published in Cambridge and Dublin Mathematical Journal [1]. He showed that when a charge 
is placed outside a grounded conducting sphere, the electric potential outside equals the 
potential of the given charge plus that of another charge imagined inside the sphere with 
its surface removed. Thus, outside the sphere, the charge 
induced on the original spherical surface has the same effect as that of the image charge 
conceived inside the sphere. The image charge is similar to the virtual image, formed in 
a mirror that would seem to emit the rays of light (to an external observer) which are originally reflected by the 
mirror. Physically, The electric field lines do not enter inside the conducting sphere (the 
electric field inside a conductor is zero [15]) like rays of light that do not enter inside a 
spherical mirror.

After Lord Kelvin, J C Maxwell applied the method to solve various electrostatic problems 
[2] involving conducting spheres and planes. His work was followed by Jeans [3] who found 
that the method was applicable to problems with dielectric boundaries as well. Due course 
of time, this method was applied to solve many complicated problems of diverse categories. 
Whereas the simplest applications are found in standard university texts (see [4] or [5]), 
some of the rigorous results can be found in [7]-[11]. The method has also been applied to 
the problems in magnetostatics [12], [13] and fluid dynamics [14] also.

However, the earlier authors in this field do not seem to have contributed on the analogy 
between the image charge and the image of a body (formed in a mirror). Probably, they did 
not take the analogy seriously. Rather, there are articles where it is indicated that no 
such analogy exists. For instance, the authors of [11] mention `in the problem of a point 
charge in the presence of a grounded perfectly conducting sphere, also a well-known example 
of an electrostatic problem, the latter does not work as a mirror'. This is not unlikely; 
because, only if a charge is placed near an infinite grounded conducting plane, the image 
charge and its position mimic the mirror image of a body formed in a plane mirror. In all 
other cases, the analogy is not at all apparent. For a charge $q$ placed at a distance $y
$ from the center of a sphere of radius $a$, the image charge $-\frac{aq}{y}$ is conceived 
at a distance $\frac{a^2}{y}$ from the center. This result does not seem to be related to 
the virtual image formation in a spherical mirror where the position and magnification of 
the image is given by a mirror equation and magnification relation [16]. Hence, it is not 
surprising that Maxwell, while defining the image charge, says `They do not correspond to 
them in actual position, or merely approximate character of optical foci' [2].

Recently, the current author has observed [18] that actually an analogy exists between the 
electrostatic image and the virtual image formed in a mirror. For the above problem, it is 
seen that all the required information about the image charge can be deduced from a mirror 
equation and a magnification formula. Instead of calculating the distances from the center 
of the sphere, if the distances are calculated from the point of intersection of the line 
joining the two charges and the sphere, the object charge distance becomes $u=(y-a)$ and 
the image charge distance becomes $v=(a-\frac{a^2}{y})$. Then, with the usual sign convention 
that for reflection in a spherical mirror, $u$ is positive and both $v$ and focal length $f$
are negative, the image charge distance can be deduced from: 
\begin{equation}
\frac{1}{u}+\frac{1}{v} = \frac{1}{f}
\end{equation}
with focal length $f = -a$. Again, the magnitude of the image charge $q'$ can be obtained 
from the following magnification formula: 
\begin{equation}
\frac{q'}{q} = -\frac{v}{u} = -\frac{a}{y}
\end{equation}

These well-known relations describing the reflection of light from a spherical mirror are 
taught in all the high school or undergraduate geometrical optics courses (see [16],[17]). 
Whereas this observation justifies the intuitive basis and the naming of image problems, 
its theoretical basis is not very clear. It is natural to ask why the equations (1.1) and 
(1.2) are valid in the image problems in electrostatics. Speculation on the reflection of 
field lines by conducting surface seems unaccountable as we do not have law of reflection 
in electrostatics as we have in optics. We find ourselves in a paradoxical situation where 
we see that our results are correct, but we cannot explain them. The present article is a 
sequel of [18] and here we shall try to give a plausible answer to this question. 

The image formation in geometrical optics by refraction through (or reflection in) lenses 
(or mirrors) is adequately described by `Fermat's principle' ([19], [20]). One can derive 
the mirror equation (1.1) for reflection in a spherical mirror, by minimizing the optical 
path length. With this in mind, we ask if an equivalent form of Fermat's Principle can be 
fitted in the existing framework of electrostatics. We propose the principle to be of the 
following form (resembling closely to Fermat's principle):
\begin{equation}
\delta\int_{\bf{a}}^{\bf{b}}{\bf{E}}\cdot d{\bf{r}}=0
\end{equation}
between two fixed points $\bf{a}$ and $\bf{b}$; here $\bf{E}$ denotes the electrostatic field. 
In this paper, we shall show that the mirror equation (1.1) in the grounded conducting sphere 
image problem can be reached starting from $\delta\int_{\bf{a}}^{\bf{b}}{\bf{E}}\cdot d{\bf{r
}}=0$ in the same manner as the usual mirror equation in optics is derived from Fermat's Principle.

The proposed principle is relevant also in the context of Green's differential equation 
[22] (known as Thomson's equation as well). This relation between the normal derivative 
of the electric field across a conducting surface and the local mean surface curvature 
is given as:
\begin{equation}
\frac{dE}{dn} = -E\left(\frac{1}{R_1}+\frac{1}{R_2}\right)
\end{equation}
-where $R_1$ and $R_2$ are the principal radii of curvature of the surface at a given location. 
There are plenty of proofs of the relation in the literature (see [23]-[26]). We shall see that 
the Euler-Lagrange equation that follow from $\delta\int_{\bf{a}}^{\bf{b}}{\bf{E}}\cdot d{\bf{r
}}=0$ can be used with Gauss's theorem to prove (1.4).

\section{Meaning of $\delta\int_{\bf a}^{\bf b}{\bf{E}}\cdot d{\bf{r}}=0$}
As the electrostatic potential difference between two fixed points along $any$ contour is the 
same, the proposition $\delta\Phi=\delta\int{\bf{E}\cdot}d{\bf{r}}=0$ appears to be redundant. 
So, what is the meaning of this statement? Although $\Phi({\bf{b}})-\Phi({\bf{a}})$ along any 
path is the same, not all of these paths are allowed by ${\bf{\nabla}}\times{\bf{E}}=\bf {0}$ 
and Laplace's equation. The proposed principle is supposed to pick up those paths that satisfy 
all these constraints. To demonstrate this claim, we derive the Euler-Lagrange equation in the 
next section.

\section{Derivation of Euler-Lagrange Equation from $\delta\int_{1}^{2}{\bf{E}}\cdot d{\bf{r}}=0$}
\subsection{Derivation}
We proposed the novel variational principle in electrostatics of the following form:
\begin{equation}
\delta\int_{1}^{2}{\bf{E}}\cdot{d\bf{r}}=0
\end{equation}
We may expand the left hand side of $(3.1)$:
\begin{eqnarray*}
 \delta\int_{1}^{2}{\bf{E}}\cdot{d\bf{r}}&=\delta\int_{1}^{2}\left[E\ dr\right]\\
                                         &=\int\left[\delta{E}dr+E\delta(dr)\right]\\
                                         &=\int\left[\left(\frac{\partial{E}}{\partial{\bf{r}}}\cdot\delta{\bf{r}}\right)dr+E\delta(dr)\right]\\
\end{eqnarray*}
The second term contains $\delta(dr)$ which can be found from Figure 1:
\begin{figure}[h]
\begin{center}
\scalebox{0.35}{\includegraphics{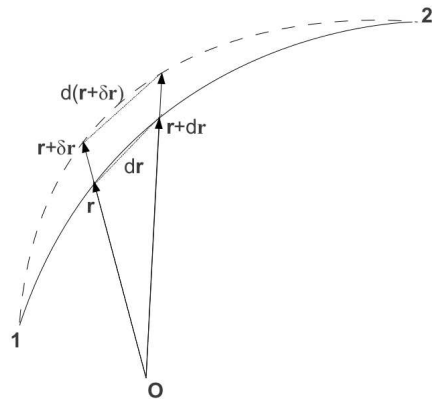}}
\caption{\label{fig:1} Variation of differential path length}
\end{center}
\end{figure}
\begin{eqnarray*}
 \delta(dr)&=|d{\bf{r}}+d({\delta\bf{r}})|-|d{\bf{r}}|\\
           &=\sqrt{d{\bf{r}}\cdot d{\bf{r}}+2d{\bf{r}}\cdot d({\delta\bf{r}})+d({\delta\bf{r}})\cdot d({\delta\bf{r}})}-\sqrt{d{\bf{r}}\cdot d{\bf{r}}}\\
           &\sim\sqrt{d{\bf{r}}\cdot d{\bf{r}}+2d{\bf{r}}\cdot d({\delta\bf{r}})}-\sqrt{d{\bf{r}}\cdot d{\bf{r}}}\\
           &=dr\left[\sqrt{1+2\frac{d{\bf{r}}\cdot d({\delta\bf{r}})}{dr^2}}\right]-dr
\end{eqnarray*}
If we take the first order term in the binomial expansion of the above square root, we get:
\begin{equation}
\delta(dr)=\frac{d{\bf{r}}}{dr}\cdot d(\delta{\bf{r}})
\end{equation}
Thus, the expression for $\delta\int_{1}^{2}{\bf{E}}\cdot{d\bf{r}}$ becomes:
\begin{eqnarray*}
 \delta\int_{1}^{2}{\bf{E}}\cdot{d\bf{r}}&=\int_{1}^{2}\left[\left(\frac{\partial{E}}{\partial{\bf{r}}}\cdot\delta{\bf{r}}\right)dr+E\delta(dr)\right]\\
                                         &=\int_{1}^{2}\left[\left(\frac{\partial{E}}{\partial{\bf{r}}}\cdot\delta{\bf{r}}\right)dr+E\left({\frac{d{\bf{r}}}{dr}\cdot d(\delta{\bf{r}})}\right)\right]\\
                                         &=\int_{1}^{2}\left[\left(\frac{\partial{E}}{\partial{\bf{r}}}\cdot\delta{\bf{r}}\right)dr-d\left(E\frac{d{\bf{r}}}{dr}\right)\cdot\delta{\bf{r}}\right]+\left[E\frac{d{\bf{r}}}{dr}\delta{\bf{r}}\right]_{1}^{2}\\
\end{eqnarray*}
-using integration by parts. The last term vanishes as $\delta{\bf{r}}=0$ at the end points. If 
we insist that $\delta\int_{1}^{2}{\bf{E}}\cdot{d\bf{r}}=0$, then the above equations imply:
\begin{eqnarray*}
 \delta\int_{1}^{2}{\bf{E}}\cdot{d\bf{r}}&=\int_{1}^{2}\left[\left(\frac{\partial{E}}{\partial{\bf{r}}}\right)dr-d\left(E\frac{d{\bf{r}}}{dr}\right)\right]\cdot\delta{\bf{r}}=0
\end{eqnarray*}
For arbitrary $\delta{\bf{r}}$, the above leads to:
\begin{equation}
 \frac{\partial{E}}{\partial{\bf{r}}}=\frac{d}{dr}\left(E\frac{d{\bf{r}}}{dr}\right)=\frac{d}{ds}\left(E\frac{d{\bf{r}}}{ds}\right)
\end{equation}

\subsection{A Digression to Validate $\delta\int_{1}^{2}{\bf{E}}\cdot{d\bf{r}}=0$}
The principle may be validated further by showing that equation (3.3) is compatible with $\nabla
\times\bf{E}=0$. Consider the x component:
\begin{equation}
\frac{\partial E}{\partial x} = \frac{dE_{x}}{ds}
\end{equation}
and similar equations for $y$ and $z$ components. If we expand the L.H.S. of (3.4), we get:
\begin{equation*}
\frac{\partial}{\partial x}\sqrt{E_{x}^2+E_{y}^2+E_{z}^2} = \frac{E_x}{E}\frac{\partial E_x}{\partial x} + \frac{E_y}{E}\frac{\partial E_y}{\partial x} + \frac{E_z}{E}\frac{\partial E_z}{\partial x}
\end{equation*}
Similarly, expanding the total derivative in the R.H.S. we get:
\begin{eqnarray*}
\frac{dE_x}{ds} & = \frac{\partial E_x}{\partial x}\frac{dx}{ds} + \frac{\partial E_x}{\partial y}\frac{dy}{ds} + \frac{\partial E_x}{\partial z}\frac{dz}{ds}\\
                & = \frac{E_x}{E}\frac{\partial E_x}{\partial x} + \frac{E_y}{E}\frac{\partial E_x}{\partial y} + \frac{E_z}{E}\frac{\partial E_x}{\partial z}
\end{eqnarray*}
-where we have used $\frac{dx}{ds}=\frac{E_x}{E}$ etc. Thus, for (3.4) to hold, we must have
\begin{equation*}
\frac{E_y}{E}\frac{\partial E_y}{\partial x} + \frac{E_z}{E}\frac{\partial E_z}{\partial x} =  \frac{E_y}{E}\frac{\partial E_x}{\partial y} + \frac{E_z}{E}\frac{\partial E_x}{\partial z}
\end{equation*}
Or, we must have (after rearranging the terms),
\begin{equation*}
\frac{E_y}{E}\left(\frac{\partial E_y}{\partial x} - \frac{\partial E_x}{\partial y}\right) =  \frac{E_z}{E}\left(\frac{\partial E_x}{\partial z} - \frac{\partial E_z}{\partial x}\right)
\end{equation*}
-which is vacuously satisfied as the factors in the parentheses in both sides of the equation are zero, 
from $\nabla\times\bf{E}=0$. Similar results follow for $y$ and $z$ counterparts of $(3.4)$ also. Thus, 
$\delta\int_{1}^{2}{\bf{E}}\cdot{d\bf{r}}=0$ is validated. We interpret the result as: although $any$ path between 
two fixed points gives the same value of the integral $\int E\ dr$, the variational principle proposed 
above allows only those paths (among all) who satisfy $\nabla\times\bf{E}=0$. Only this $\bf E$ derives 
from the unique solution to Poisson's or Laplace's equation.

\section{Mirror Equation from Variational Principle}
Before attacking the problem in electrostatics, let us review the corresponding problem in optics. The 
``Fermat's principle'' is the variational principle used in optics. It is given as [20] $\delta\int_{1
}^{2}{\bf{k}}\cdot{d\bf{l}}=0$ (here $\bf{k}$ is the wave vector).
\subsection{In Optics}
\begin{figure}[h]
\begin{center}
\scalebox{0.50}{\includegraphics{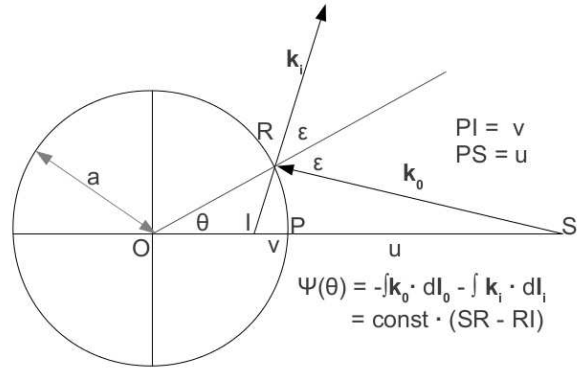}}
\caption{\label{fig:1} Mirror Equation in Optics}
\end{center}
\end{figure}
In Figure 2, a ray of light ($\bf{k_o}$) from source $S$ gets reflected from a point $R$ on the 
mirror surface. To an observer, light ray ($\bf k_i$) seems to come from the image $I$. In this 
system, phase $\psi$ is given as:
\begin{eqnarray*}
 \psi& =\int_{S}^{R}{\bf{k_{o}}}\cdot\ d{\bf{l_{o}}}+\int_{R}^{I}{\bf{k_{i}}}\cdot\ d{\bf{l_{i}}}\\
     & =\int_{S}^{R}\ k_{o}\ dl_{o}-\int_{R}^{I}\ k_{i}\ dl_{i}\\
     & =constant\times\left[\int_{S}^{R}\ dl_{o}-\int_{R}^{I}\ dl_{i}\right]
\end{eqnarray*}
The subscripts $_o$ and $_i$ denote the outside and inside of the spherical mirror, respectively.
The negative sign comes before $k_i$ because its actual direction is opposite to that of $d\bf l
_i$. Thus, $\ optical\ path\ length$ $L_{op}$ is given as (in the small angle limit):
\begin{equation}
\ L_{op}=SR-RI=u-v+\frac{1}{2}\ a^2\left(\frac{1}{u}-\frac{1}{v}+\frac{2}{a}\right)\theta^2
\end{equation}
The symbols are explained in Figure 2. We know that the variation of the optical path with respect 
to the angle $\theta$ gives the desired mirror formula (1.1) with focal length $|f|=\frac{a}{2}$. 
The minimum $L_ {op}$ is guaranteed by the condition $\frac{\partial L_{op}}{\partial\theta}|_{\theta\rightarrow0}=0$. 

\subsection{In Electrostatics}
It has been found [18] that the location of the image charge in the grounded conducting sphere 
image problem can be extracted from a mirror equation (1.1). Whereas the distance of the image 
charge from the center of the sphere can also be found from standard texts ([4]-[5]), we shall 
make use of the principle $\delta\int_{\bf{a}}^{\bf{b}}{\bf{E}}\cdot d{\bf{l}}=0$ to see if we 
can derived (1.1). Instead of a point charge, we assume that the real charge is distributed 
uniformly over a very small sphere $S$ which is finite nevertheless. We call its center as $1$. 
If we do not assume this, $\int _1^{R}{\bf{E_o}}\cdot d{\bf{l_o}}$ will diverge from the lower 
limit. Since inversion of a sphere in a bigger sphere is another sphere [21], the image charge 
$I$ is also spherical. We denote its center by $2$. Evidently, $\int_{R}^{2}{\bf{E_{i}}}\cdot 
d{\bf{l_{i}}}$ does not diverge from the upper limit (potential at a point inside a continuous 
charge distribution is finite). 

Let us assume that we have no prior information about the position or the value of image charge. 
We define the $electric\ path\ potential$ (refer to Discussions) between the real charge and the 
image charge as (Figure 3): 
\begin{figure}[h]
\begin{center}
\scalebox{0.50}{\includegraphics{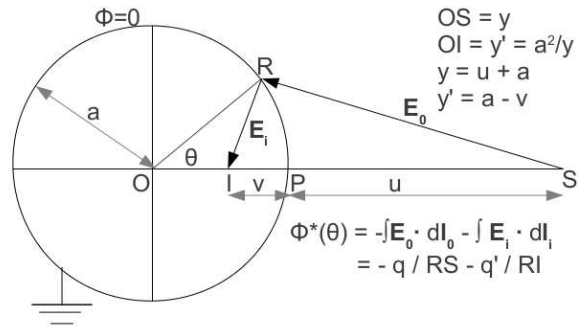}}
\caption{\label{fig:1} Mirror Equation in Image Problem}
\end{center}
\end{figure}

\begin{equation}
 \Phi^{*}=-\int_{1}^{R}{\bf{E_{o}}}\cdot d{\bf{l_{o}}}-\int_{R}^{2}{\bf{E_{i}}}\cdot d{\bf{l_{i}}}=-\int_{1}^{R}{\bf{E_{o}}}\cdot d{\bf{l_{o}}}+\int_{2}^{R}{\bf{E_{i}}}\cdot d{\bf{l_{i}}}
\end{equation}
The symbols are explained in figure 3. Notice that the $^{*}$ sign is applied to make it 
explicit that $\Phi^{*}$ between $1$ and $2$ is different from the potential $\Phi(\bf{r})$ 
which is the solution of Laplace's equation outside the sphere. Thus,
\begin{equation}
 \Phi^{*}=[\Phi_{S}(R)-\Phi_{S}(1)]-[\Phi_{I}(R)-\Phi_{I}(2)]
\end{equation}
or, more explicitly,
\begin{equation}
 \Phi^{*}=\frac{q}{\sqrt{y^2+a^2-2ya cos\theta}}-\frac{q'}{\sqrt{a^2+y'^2-2ay' cos\theta}}-C
\end{equation}

-where $C = (\Phi_{S}(1)-\Phi_{I}(2))$ is finite and independent of $\theta$. At this point we do 
not specify the potential on the sphere. This boundary condition will be invoked later. Writing $
-2ya\ cos\theta=-2ya+2ya(1-cos\theta)$, we get:
\begin{equation}
 \fl{\Phi^{*}=\frac{q}{\sqrt{y^2+a^2-2ya+2ya(1-cos\theta)}}-\frac{q'}{\sqrt{a^2+y'^2-2ay'+2ay'(1-cos\theta)}}-C}
\end{equation}
Using real charge distance $u=(y-a)$ and image charge distance $v=(a-y')$, we find
\begin{equation}
 \fl{\Phi^{*}=\frac{q}{\sqrt{u^2+2a(a+u)(1-cos\theta)}}-\frac{q'}{\sqrt{v^2+2a(a-v)(1-cos\theta)}}-C}
\end{equation}
Now, we set the derivative of $\Phi^*$ with respect to the angle $\theta$ to zero:
\begin{equation}
 \fl{\frac{\partial\Phi^{*}}{\partial\theta}=-a\ sin\theta\left[\frac{q(a+u)}{(u^2+2a(a+u)(1-cos\theta))^{\frac{3}{2}}}-\frac{q'(a-v)}{(v^2+2a(a-v)(1-cos\theta))^{\frac{3}{2}}}\right]=0}
\end{equation}
We are following the notion that electric path potential $\Phi^{*}$ must be stationary along all the 
electric field paths. This must be the case because field lines from all possible angles are responsible 
for image formation. Thus, $\Phi^{*}$ is independent of the parameter $\theta$. No extreme condition like 
$\frac{\partial\Phi^*}{\partial\theta}|_{\theta\rightarrow 0}$ is needed. 
Let us now use the boundary condition (1.2) that at the pole $P$, the potential $\Phi(P)=0$:
\begin{equation}
 \frac{q}{u}=-\frac{q'}{v}
\end{equation}
which can be used to assert that $q=uk$ and $q'=-vk$ where $k$ is a non-zero constant. Using (4.8), 
(4.7) reduces to
\begin{equation}
 \fl{-ka\ sin\theta\ \left[\frac{u(a+u)}{(u^2+2a(a+u)(1-cos\theta))^{\frac{3}{2}}}+\frac{v(a-v)}{(v^2+2a(a-v)(1-cos\theta))^{\frac{3}{2}}}\right]=0}
\end{equation}
In general $\theta\neq 0$; therefore, we have,
\begin{equation}
 {\left[\frac{u(a+u)}{(u^2+2a(a+u)(1-cos\theta))^{\frac{3}{2}}}+\frac{v(a-v)}{(v^2+2a(a-v)(1-cos\theta))^{\frac{3}{2}}}\right]=0}
\end{equation}
From (4.10), we get the following:
\begin{equation}
 \fl{u^2(a+u)^2[v^2+2a(a-v)(1-cos\theta)]^3-v^2(a-v)^2[u^2+2a(a+u)(1-cos\theta)]^3=0}
\end{equation}
From the problem, it is clear that (4.11) must hold for all values of $\theta$. It is an 
identity rather than an equation. So, each coefficient of $(1-cos\theta)$ or its higher 
powers must be equated to zero. The term independent of $\theta$ yields:
\begin{equation}
 u^2(a+u)^2v^6=v^2(a-v)^2u^6
\end{equation}
Canceling common non-zero factors and taking positive square root, we get
\begin{equation}
 u^2(a-v)=v^2(a+u)
\end{equation}
This can be written as
\begin{equation}
 a(v+u)(v-u)=-uv(u+v)
\end{equation}
Canceling the common factor $(u+v)$, we get
\begin{equation}
 a(v-u)=-uv
\end{equation}
Or, in more familiar form,
\begin{equation}
 \frac{1}{u}-\frac{1}{v}=-\frac{1}{a}
\end{equation}
The reader is strongly encouraged to check that the same or trivial result follows 
by equating higher powers of $(1-cos\theta)$. Now, we know that real charge distance is $u=y
-a$. Then, (4.16) gives the image charge distance $v=\frac{a}{y}(y-a)$ behind the mirror. The 
boundary condition (4.8) gives the magnitude of image charge $q'=-\frac{aq}{y}$. After we get 
information about both distance and value of the image charge, we can construct the Green's 
function for the problem and the problem is solved.

Taking the sign convention, real charge distance $u$ to be positive when image charge distance $v$ 
is negative and the focal length $a$ is also negative, (4.16) can be put in the familiar form:

\begin{equation}
 \frac{1}{u}+\frac{1}{v}=\frac{1}{f}
\end{equation}

\section{Proof of Green's Differential Equation}
In vector notation, the ``Euler-Lagrange'' equation of $\delta\int_{\bf{a}}^{\bf{b}}{\bf{E}}\cdot d{\bf
{l}}=0$ is given by:
\begin{equation*}
{\bf{\nabla}}|{\bf{E}}|=\frac{d}{ds}({\bf{|E|}}\frac{d{\bf{r}}}{ds})=\frac{d{\bf{E}}}{ds}
\end{equation*}

Let us write the equation for an electric field near a conducting surface. The electric field 
is perpendicular to the conducting surface [15]. That is, $\bf{E}=\bf{E}^{||}+\bf{E}^{\bot}=\bf
{E}^{\bot}$ where $\bf{E}^{\bot}$ and $\bf{E}^{||}$ denote normal and tangential components of 
the electric field. Denoting infinitesimal displacement $ds$ along direction of $\bf E^{\bot}$ 
by $dn$ and the unit vector normal to the conducting surface by $\bf{\hat{n}}$, we have:
\begin{eqnarray*}
{\bf{\nabla}}\cdot{\bf{E}} & = {\bf{\nabla}}\cdot({|\bf{E}|\hat{n}})\\
                           & = {\bf{\nabla}}{|\bf{E}|}\cdot{\bf{\hat{n}}} + |\bf{E}| {\bf{\nabla}}\cdot{\bf{\hat{n}}}\\
                           & = \frac{d{\bf{E}}}{dn}\cdot{\bf{\hat{n}}} + 2\kappa|\bf{E}|
\end{eqnarray*}
-where we have used mean curvature, $\kappa=\frac{1}{2}\nabla\cdot\bf{\hat{n}}$ [27]. If we work 
just outside the conductor, Laplace's equation holds and ${\bf{\nabla}}\cdot{\bf{E}}=0$. Thus, in 
this region, 
\begin{eqnarray*}
\frac{d{|\bf{E}|}}{dn} & = \frac{d{\bf{E}}}{dn}\cdot{\bf{\hat{n}}}\\
                       & = -2\kappa|\bf{E}|\\
                       & = -|{\bf{E}}|\left(\frac{1}{R_1}+\frac{1}{R_2}\right)
\end{eqnarray*}
Thus, with little effort we have proved Thomson's equation:
\begin{equation}
\frac{dE}{dn} = - 2\kappa E = -E \left(\frac{1}{R_1}+\frac{1}{R_2}\right)
\end{equation}

In this proof, we see the value of the vector relation $\nabla{E}\cdot{d{\bf{n}}}=dE$, which we 
identified with the ``Euler-Lagrange'' equation of $\delta\int_{\bf{a}}^{\bf{b}}{\bf{E}}\cdot d
{\bf{l}}=0$. 

\section{Discussions}
In this article, we saw that a novel variational principle can be conceived in electrostatics. 
We used this principle to justify the observations made in [18] and to prove Green's 
differential equation [22]. The `Euler-Lagrange' equation obtained from this principle was 
identified to be a well-known vector relation.

In the proof of the mirror equation, the `electric path potential' was constructed in a way so that 
the contour of the integral $\int{\bf{E}}\cdot d\bf{l}$ stays always superimposed with the local 
$\bf E$ field direction. Exactly the same is done in optics (subsection 4.1). The contributions 
to the phase from wave vectors $\bf k_o$ and $\bf k_i$ are treated separately. So, electrostatic 
path potential is just the electrostatic twin of the optical path length. In this proof, we did 
not use any `law of reflection' which, in optics, is actually a manifestation of the boundary conditions of the 
wave vector $\bf{k}$ at the interface between two media. Our proof was facilitated by the use of (4.8), which is 
the boundary condition in the context of this problem. Thus, the boundary conditions play a very 
crucial role in the formation of $all$ virtual images. An ideal point charge could not be taken 
for this work, to avoid divergence. That is not a big problem, as no charge is ideal in reality. 
Also, the concise proof of Green's differential equation shows the relevance of the `Euler-
Lagrange' equation.

Image method is also applicable to magnetostatic boundary value problems (for example, one can 
solve the problem of a magnetic dipole placed in front of an infinite superconducting plane by 
image method). Other problems may be found in various literatures (Hague [12] or Q.G.Lin [13]).
They can be seen in the light of the present article in the following way. In the current free 
region, magnetic field may be expressed as the gradient of a scalar potential $U(\bf{r})$, and 
both $\nabla\times{\bf{B}}=\bf0$ and Laplace's equation $\nabla^2 U({\bf r})=0$ apply. Thus, we 
see that a calculation parallel to the one described in section $3$ (with $U(\bf{r})\Leftrightarrow
\Phi(\bf{r})$) is possible if $d\bf s$ is taken parallel to $\bf{B}$ field lines. Then, we can 
speak of $\delta\int_{\bf{a}}^{\bf{b}}{\bf{B}}\cdot d{\bf l}=0$ in magnetostatics also. Physically, 
magnetic field cannot penetrate inside a superconductor and the interface again behaves like a 
mirror that effectively reflects magnetic field lines. The argument is strengthened by the 
observation that Green's differential equation (section $5$) is valid for magnetostatic field as 
well [28].

Clearly, the work presented in this paper reveals an unfamiliar face of the image problems. We 
feel that the proposed principle may be explored even further to understand the physical world. 
Hopefully, the interdisciplinary nature of the article will attract the general audience.

\section{Acknowledgements}
I thank Dr. Debapriyo Syam for his encouragements. I also thank my friends Swastik, 
Tanmay, Manoneeta and Tamali for precious discussions with them.

\end{document}